\documentclass[aps,prb,reprint,a4paper,superscriptaddress,showpacs,showkeys]{revtex4-1}
\usepackage{graphicx,hyperref,mathptmx}

\begin{document}

\preprint{APS/123-QED}
\title{Local Magnetism and Spin Dynamics of the Frustrated Honeycomb Rhodate
Li$_{2}$RhO$_{3}$}
\date{\today}
\author{P. Khuntia}
\altaffiliation[Present address: ]{Department of Physics, Indian Institute of Technology, Madras Chennai-600036, India}
\affiliation{Max Planck Institute for Chemical Physics of Solids, 01187 Dresden, Germany}
\author{S. Manni}
\altaffiliation[Present address: ]{Department of Physics and Astronomy, Iowa State University, Ames, Iowa, USA}
\affiliation{EP VI, Center for Electronic Correlations and Magnetism, Augsburg
University, D-86159 Augsburg, Germany}
\author{F. R. Foronda}
\affiliation{Oxford University Department of Physics, Clarendon Laboratory,
Parks Road, Oxford OX1 3PU, United Kingdom}
\author{T. Lancaster}
\affiliation{Durham University, Centre for Materials Physics, South Road, Durham, DH1
3LE, United Kingdom}
\author{S. J. Blundell}
\affiliation{Oxford University Department of Physics, Clarendon Laboratory,
Parks Road, Oxford OX1 3PU, United Kingdom}
\author{P. Gegenwart}
\affiliation{EP VI, Center for Electronic Correlations and Magnetism, Augsburg
University, D-86159 Augsburg, Germany}
\author{M. Baenitz}
\thanks {corresponding author: baenitz@cpfs.mpg.de}
\affiliation{Max Planck Institute for Chemical Physics of Solids, 01187 Dresden, Germany}

\begin{abstract}
We report magnetization, heat capacity, $^{7}$Li nuclear magnetic
resonance (NMR) and muon-spin rotation ($\mu$SR) measurements
on the honeycomb 4d$^{5}$ spin liquid candidate Li$_{2}$RhO$_{3}$. The
magnetization  in small magnetic fields provides evidence for a
partial spin freezing of a small fraction of Rh$^{4+}$ moments at 6~K whereas the
Curie-Weiss behavior  above 100~K suggests a pseudo-spin--1/2 paramagnet with a moment of about 2.2~$\mu _{B}$. The
magnetic specific heat ($C_{\rm m}$) exhibits no field dependence and
demonstrates the absence of long range magnetic order down to 0.35
K. $C_{\rm m} / T$ passes through a broad
maximum at about 10 K and $C_{\rm m}\propto T^{2}$ at low temperature.
Measurements of the spin-lattice relaxation
rate ($1/T_{1}$) reveal a gapless slowing down of spin fluctuations on
cooling with $1/T_{1}\sim T^{2.2}$.
The results from NMR and
$\mu$SR are consistent with a scenario in which a minority of Rh$^{4+}$
moments are in a short-range correlated frozen state and coexist with
a majority of moments in a liquid-like state that continue to fluctuate at low
temperature.
\end{abstract}

\maketitle

\section{Introduction}

The physics of $S=\frac{1}{2}$ quantum magnets (QM) is extremely rich, owing
to the variety of magnetic exchange interaction networks in
different systems, as determined by the lattice geometry and the
orbital hybridization \cite{LB,SS}.  Systems studied so far include quasi
1D-linear chains, planar 2D-systems (ladders, kagome-layers, triangles
or square lattices) or more complex 3D-structures such as the
hyperkagome or pyrochlore lattices. Recently, the field of $S=\frac{1}{2}$
quantum magnetism has been extended away from 3d ions (such as
those containing Cu$^{2+}$ or V$^{4+}$ ions) towards 4d and 5d systems
\cite{LB}. In these materials, an effective $j_{\rm eff}=\frac{1}{2}$ moment
can be realized due to strong spin orbit
coupling (SOC) and in certain compounds the presence of frustration
is suspected to lead
to a quantum spin liquid (QSL) ground state \cite{CL}. In general,
having the energy of the SOC, the Coulomb interaction (parameterised
by $U$) and the crystal electric field splitting (CEF) of the same
order of magnitude leads to highly degenerate magnetic states and
complex excitations for many 4d and 4d QMs. These excitations can be
gapless or gapped, but their nature is complicated by the presence of
disorder and anisotropic interactions, and their understanding is
hindered by the scarcity of
model materials\cite{LB,CL}.
One approach to describe the highly
degenerate states in such frustrated systems utilizes a fermionic band-like picture with chargeless
spinon ($S=\frac{1}{2}$) excitations. The fermionic spinon concept was first
introduced in cuprates \cite{JTK} and for organic Cu-based QSLs
\cite{SL,RK}. It was later established in other QMs, such as spin
chains\cite{MM}, 2D-systems \cite{JN1,TH, JS} and 3D-networks
including pyrochlore \ lattices\cite{JSG,BN}.

The honeycomb 4d and 5d
planar QMs have become attractive systems to study following the discovery of
graphene, a honeycomb, 2D Dirac semimetal (SM); its unique properties
stem from its linear dispersive modes ($E\sim k$) and its $T$-linear
density of states at the Fermi level $N(E_{F})$ \cite{AKN}.  These
properties lead to a
$T^2$-behavior in the electronic specific
heat ($\propto T N(E_{\rm F})$) and a $T^3$-power
law for the spin-lattice relaxation rate $1/T_{1}$
($\propto T N^2(E_{\rm F})$) \cite{AKN,BD}.
By analogy 2D-honeycomb QSLs, with linear
dispersive fermionic spinon bands and low energy gapless magnetic
excitations (spinons or Majorana fermions), are expected to exhibit
a $T^2$-behavior
in the magnetic specific heat \cite{RM1,RM3, DQSL} and also
power-law spin-lattice relaxation: $1/T_1\sim T^n$
\cite{DQSL,RM1,RM3,JY,XY,JK}.

The theoretically-solvable Heisenberg-Kitaev model
\cite{JC,AK,ST1,JC2} predicts that a honeycomb lattice decorated with
$j_{\rm eff}=\frac{1}{2}$ pseudospins can have a QSL ground state.
Experimental
realizations of this include (Li,Na)$_{2}$IrO$_{3}$, and
$\alpha$-RuCl$_{3}$~\cite{Williams, AB,
  SHB,NP1,ST,SKC,RT,MB,TT,SM1,SM2,ST2}. Na$_{2}$IrO$_{3}$ displays
zig-zag magnetic ordering, while $\alpha$-Li$_{2}$IrO$_{3}$ exhibits
incommensurate spiral ordering. $\alpha$-RuCl$_{3}$ exhibits a complex
magnetic ordered state while recent NMR results suggest a gapping out
of magnetic excitations towards low temperatures once the order is
suppressed by magnetic field~\cite{SHB,MB}. Surprisingly, the
structural 4d-homologue Li$_{2}$RhO$_{3}$ also shows insulating
behavior in spite of reduced spin-orbit interactions \cite{IIM} and
even more interestingly this system exhibits no sign of long range
magnetic ordering (LRO), unlike its Ir counterpart
\cite{YKL}. Magnetic exchange between Rh$^{4+}$-ions are expected to
be highly frustrated, which makes this pseudospin $j_{\rm eff}=\frac{1}{2}$
system a promising candidate for a Kitaev quantum spin liquid. Here we
provide a comprehensive account of the local magnetic properties of
Li$_{2}$RhO$_{3}$ probed by $^{7}$Li ($I=3/2$) NMR and $\mu$SR
accompanied by magnetization and heat capacity measurements down to
0.4 K.

\section{Results and Discussion}

Polycrystalline samples of Li$_{2}$RhO$_{3}$ were synthesized by a method
described elsewhere (see the supplemental information~\cite{SM}). Shown in Fig. 1(a) is the
temperature dependence of the \textit{d.c.}-magnetic susceptibility $\chi (T)$
measured following zero-field cooled (ZFC) and field-cooled (FC) protocols at
10~mT. The ZFC and the FC $\chi (T)$  curves split at 6~K, which may arise from short range magnetic order (SRO) due to
a partial freezing of Rh-moments. However, the splitting in $\chi (T)$ is small, which indicates
that probably only a small fraction of moments participate in the glassy state. SRO effects (``spin
freezing'') admixed onto the quantum spin liquid state have been discussed in
quite a large number of materials, such as Na$_{4}$Ir$_{3}$O$_{8}$ and
Ni$_{2}$Ga$_{2}$S$_{4}$ \cite{SN,muSR,YQM,SN1,YO,RD,ACS,YJU,HDZ,PK1}. The
Curie-Weiss (CW) fit (Fig.~1b) in the  range $100\leq T \leq 300$~K
yields an effective moment of 2.2 $\mu _{B}$ per Rh-ion. This is well above the
spin-only value for the $S=\frac{1}{2}$ low-spin configuration of the
4d$^{5}$-state of Rh$^{4+}$, which points towards a moderate spin orbit coupling \cite{YL,IIM,VT}. The negative
sign of the Curie Weiss temperature $\theta_{\rm CW} = -60$~K suggests
the prominence of antiferromagnetic (AFM) correlations between
Rh$^{4+}$-moments. The exchange interaction between the nearest
neighbor Rh$^{4+}$-moments can be determined from
the high temperature series expansion
(HTSE) frequently used for honeycomb lattices with moderate SOC (such
as 4d$^{5}$ Ru$^{3+}$ ions in  $\alpha$-RuCl$_{3}$).
 The HTSE yields an AFM interaction of $J/k_{\rm B}\approx 75\pm 5$~K and is in reasonable agreement
with that obtained from the mean field approximation (MFA)~\cite{SM,HTSE1}. The 
a.c.-susceptibility exhibits a peak at about $T_{\rm g}\approx 6$~K~\cite{SM} and the peak positions are weakly frequency
dependent, showing the role of dissipative spin dynamics in driving such a
short range spin freezing mechanism. The origin of this partial spin
freezing might be related to the presence of local disorder in the
lattice of
Li$_{2}$RhO$_{3}$~\cite{YS1,YS2,VT,PGT}  (see discussion in \cite{SM})  and the glassy feature smears out at
higher fields ($\mu_0H>1$~T)(see Fig. 1b). It may be noted that the
spin freezing effect on the magnetization and the heat capacity in Li$_{2}$RhO$_{3}$ is rather minor in comparison with
the textbook spin glass materials~\cite{JAM}.

\begin{figure}
\includegraphics[width=\columnwidth]{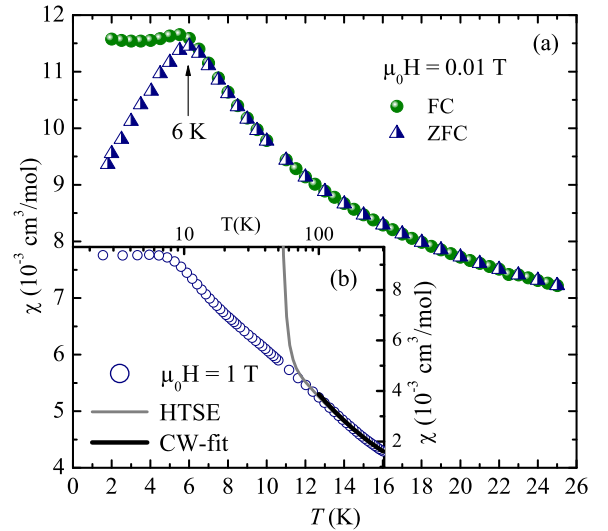}
\caption{(Color online) (a) The
temperature dependence of the ZFC and FC susceptibility $\chi (T)$
measured in an applied field of 0.01~T. (b) $\chi(T)$ at 1~T, with HTSE (high temperature series expansion)  and CW ( Curie-Weiss) fits as discussed in
the text. }
\end{figure}

\begin{figure}
\includegraphics[width=\columnwidth]{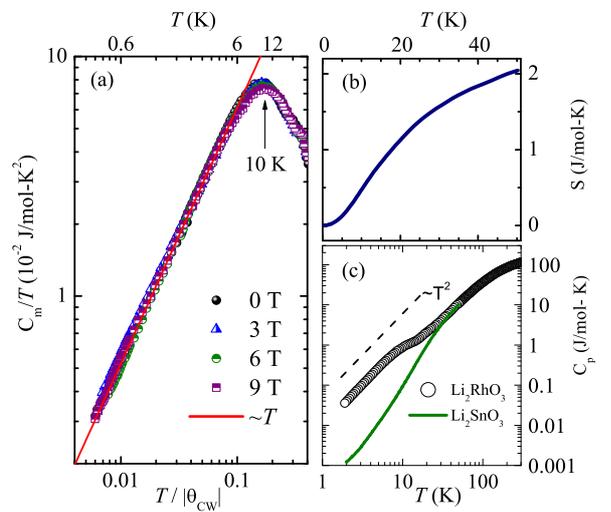}
\caption{(Color online)  (a) Magnetic heat capacity 
co-efficient ($C_{\rm m}/T$) in various fields as a function of
$T/|\theta_{\rm CW}|$.  The upper axis shows the absolute
$T$-dependence and the solid line represents a $T$-linear fit,
as discussed in the text. (b)
$T$-dependence of the magnetic entropy in zero field. (c)
$T$-dependence of the total heat capacity in zero field for
Li$_{2}$RhO$_{3}$ compared with the non-magnetic homologue
Li$_{2}$SnO$_{3}$.  The dashed line indicates a  \textit{T}$^{2}$-behaviour. }
\end{figure}

The heat capacity coefficient $C_{\rm m}/T$ obtained in different
magnetic fields is shown in Fig.~2(a).
The heat capacity exhibits no signature of LRO down to 0.35 K. The
magnetic heat capacity ($C_{\rm m}$) 
was obtained by subtracting the lattice contribution using Li$_{2}$SnO$_{3}$%
[see Fig.~2(c)] as a reference. As shown in Fig.~2(a), $C_{\rm m}/T$
displays a broad maximum at about 10~K, which could be associated
with the highly frustrated nature of the system as discussed in spin liquids 
\cite{LB,YO,JGC}. The strength of the exchange coupling and dimensionality
of the system accounts for the position of the broad maximum in
$C_{m}$ and it varies as $T/|\theta_{\rm CW}|$ in frustrated magnets. In
Li$_{2}$RhO$_{3}$, we found $T/|\theta_{\rm CW}|\approx 0.16$, which is
comparable with those values in other 3d and 5d frustrated
magnets \cite{AP,SN,YO}. The magnetic entropy $S_{\rm m}=\int C_{\rm
  m}/T\,{\rm d}T$ up to 45~K was found to be only 35\% [$\approx
2.04$~J/mol\,K,  Fig.~2(b)] of $R\ln 2$ ($\approx 5.76$~J/mol\,K), consistent with the presence of short-range spin
correlations. Below 10~K, $C_{\rm m}$ exhibits a $T^2$-behavior
[Fig.~2(a)]
indicating
the persistence of spin dynamics with low lying gapless excitations, which
is in agreement with the finite value of $\chi$ at low $T$ in the
context of the QSL state. The $T^{2}$-dependence of $C_{\rm m}$
is frequently found in 4d and 5d quantum magnets as a
fingerprint of the spin liquid ground  state \cite{Th1,ST1,Th3}.

\begin{figure}
\includegraphics[width=\columnwidth]{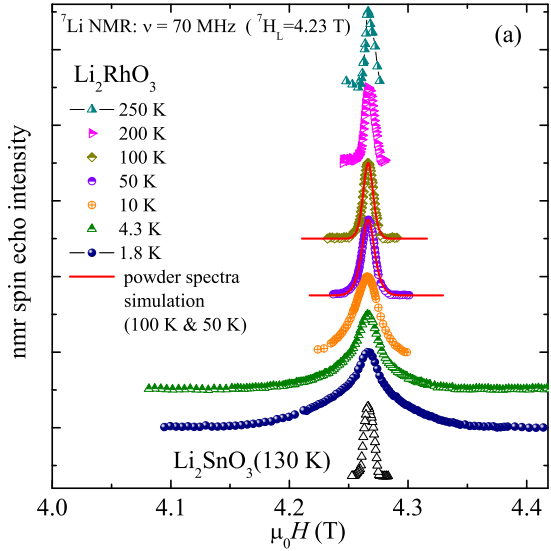}
\caption{(Color online) Representative  field-swept $^{7}$Li-NMR\ spectra in Li$_{2}$RhO$_{3}$ at different temperatures (the
solid line is a simulation for 100~K and for 50~K). At the bottom 
of the figure we show the $^{7}$Li-NMR spectrum at 130~K
for the non-magnetic structural homologue Li$_{2}$SnO$_{3}$. }
\end{figure}

The $^{7}$Li NMR powder spectra at 70 MHz shown in Fig.~3(a) show a
single
Li-NMR line which exhibits a clear
broadening towards low temperatures without any strong anisotropy. The
spectra consist of superimposed intensities from three powder averaged
Li-lines from the three Li-sites in the lattice structure (see
Supplementary Material \cite{SM}
for more details).  The inset
of Fig.~4(a) represents the $T$-dependent NMR shift, $K(T)$, 
estimated from the simulation of each powder spectrum [see solid
line in Fig.~3(a)]. The $T$-dependence of the shift is reminiscent of
the bulk susceptibility [Fig.~1(b)]. $K(T)$ consists of a 
$T$-dependent part $K_{\rm Rh}(T)$ due to the coupling of
the Rh$^{4+}$ moments with the Li nuclear spins and a nearly a 
$T$-independent orbital part $K_{\rm orb}$ which is enhanced due to the presence of moderate spin-orbit interaction \cite{Th3,BY}.
The linear scaling between $K$ and $\chi$ 
(given by $K=A_{\rm hf}\chi/N_{\rm A}$) at high-$T$ yields a hyperfine coupling
constant $A_{\rm hf}=-(0.3\pm 0.06)$~kOe/$\mu _{\rm B}$ between
the $^{7}$Li
nucleus and the Rh$^{4+}$ electron spin. Fig.~4(a) shows the NMR
linewidth
[full width at half maximum (FWHM)] divided by the resonance field (therefore relative linewidth,
$\delta H=\Delta H/H$) at two NMR frequencies.
The relative linewidth $\delta H$ exhibits no field dependency and follows
the bulk susceptibility [see Fig.~1(b)]. To account for the effect of
the first order quadrupolar splitting on the line broadening of the
$^{7}$Li NMR powder spectra we have investigated the non-magnetic
homologue
Li$_{2}$SnO$_{3}$ under the same NMR conditions (see Supplemental
Material \cite{SM}). This gives
clear evidence that the low temperature broadening is a generic
feature of
Li$_{2}$RhO$_{3}$ and the scaling with the bulk susceptibility demonstrates
the magnetic origin of the broadening. The broadening is associated
with static and slow fluctuating hyperfine field contributions at the nuclei
sites. It is remarkable that $\delta H$ is independent of magnetic field.
This implies that the absolute width is field dependent, suggesting that
at these fields the system is not yet in the fully polarized state and
sizeable correlations among Rh$^{4+}$ moments are still present. This is
consistent with the absence of LRO in $C_{\rm m}$ down to 0.35~K.
The saturation of $\delta H(T)$
at low $T$ indicates the persistence of a quasi-static distribution of
local magnetic fields and a slowing down of magnetic fluctuations such that
Rh$^{4+}$-moments fluctuate with a frequency less than the NMR frequency. The
fact that above approximately 100~K [see Fig.~4(a) and Supplementary Material~\cite{SM}] Li$_{2}$RhO$_{3}$ and
Li$_{2}$SnO$_{3}$ have comparable NMR linewidths, suggests that the effect of anti site order (Li--Rh or Li--Sn) discussed frequently in the literature \cite{VM} in the linewidth could be neglected
(see discussion in \cite{SM}). The magnetic moment of 0.8~$\mu _{B}$  estimated from the NMR linewidth at about 4~K and in a NMR field of 4.23~T (70~MHz) is small compared to the Rh$^{4+}$ Curie-Weiss moment and suggests the presence of strong quantum fluctuations induced by
magnetic frustration \cite{YS,YQM}. We found no loss of
NMR signal intensity, typical for some disordered materials, which indicates
that Li$_{2}$RhO$_{3}$ is not a conventional spin glass material. This
scenario is further supported by the absence of rectangular shaped powder
averaged NMR spectra expected for materials that show LRO \cite{YY} and moreover by
the field independence of the relative linewidth. The quasi-static NMR results presented so far support the scenario of a minority part
of moments in a short range like frozen state coexisting with a majority of
moments which remain liquid-like and which fluctuate at low-$T$
\cite{SN,muSR,YQM,SN1,YO}.

\begin{figure}
\includegraphics[width=\columnwidth]{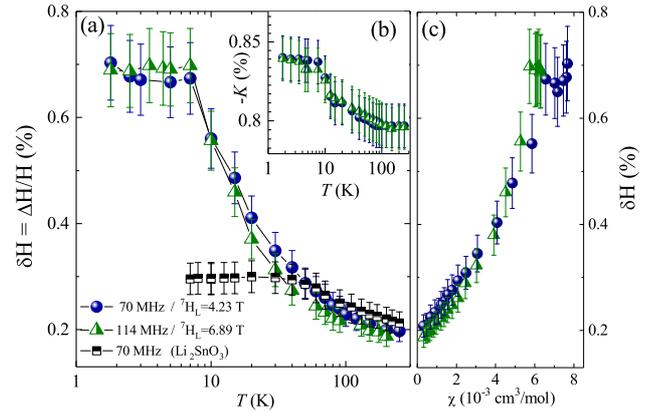}
\caption{(Color online) (a) The $T$-dependence of the full width at half
  maximum (FWHM) linewidth divided by the resonance field (=relative
  linewidth, $\delta H = \Delta H/H$) at 70~MHz and 114~MHz
  compared with $\delta(H)$
of the non-magnetic homologue Li$_{2}$SnO$_{3}$.  (b) $T$-dependent NMR
shift, $K$, at 70~MHz and 114~MHz. (c) $\delta H(T)$  vs. $\chi(T)$
(obtained at the NMR fields) with $T$ as an
implicit parameter. The  Larmor fields are calculated by using the
$^7$Li gyromagnetic ratio of 16.5459~MHz/T.}
\end{figure}

\begin{figure}
\includegraphics[width=\columnwidth]{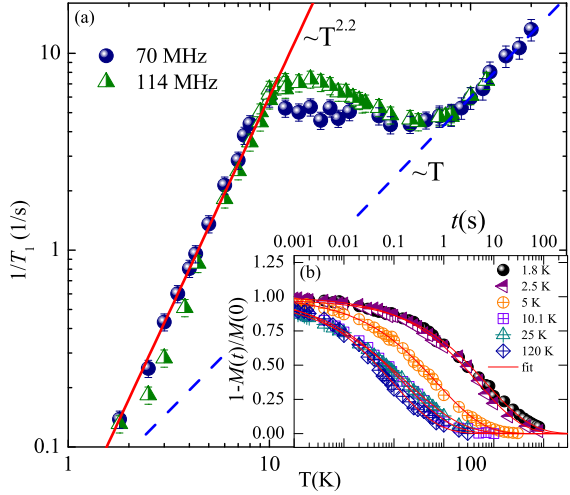}
\caption{(Color online) (a) The $T$-dependence of $1/T_1$
  at two NMR frequencies (fields). The solid
line indicates a $T^{2.2}$-behavior below 10~K wheras the dashed
line depicts the $T$-linear behavior around 100~K. (b) Longitudinal
magnetization recovery curves $M(t)$ at various temperatures in a
semi-log
plot. The solid curved lines are the
individual fits of the data with a stretched exponential function (see
Supplemental Materials \cite{SM}
for more details) at various temperatures.}
\end{figure}

NMR spin-lattice relaxation rate measurements are very suitable to probe
slow spin excitations because in general $1/T_1$ tracks the 
$q$-dependent complex dynamic spin susceptibility (see \cite{SM}
for more details). Fig.~5(a) shows $1/T_1$ vs.\ $T$ at two
different NMR frequencies (fields). Towards low temperatures $1/T_1$
decreases linearly with $T$ and passes through a broad maximum
around 10~K. This maximum could not be associated with
a conventional SG freezing
where a critical slowing down of spin fluctuations at $T_{\rm g}$
leads
to a very short $T_1$
and an NMR signal wipeout at low and intermediate $T$
\cite{YQM,YS}. As shown in Fig.~5(a), $1/T_1$
decreases on further cooling below 10~K and displays a pronounced
$T^{2.2}$-behavior down to 1.8~K. In principle, $1/T_{1}$ tracks the spectral
density of the Fourier transform of the time correlation function of the
transverse component $\delta h_{+}$ of the fluctuating local field at
nuclear sites $h_{\pm }(0)$ with the nuclear Larmor frequency
$\omega_{\rm N} = \gamma H$ as \cite{TM1,BPP,Abragam}
$\frac{1}{T_{1}}=\frac{\gamma_{\rm N}^{2}}{2}\int_{-\infty}^{+\infty}\langle
h_{\pm }(t)h_{\pm}(0)\rangle {\rm e}^{i\omega_{\rm N}t}\,{\rm d}t$, where
  $\gamma_{\rm N}$ is the gyromagnetic ratio of the nuclear spin.
  Assuming the
  time correlation function varies as e$^{-\Gamma t}$, one can express
  $R=\frac{1}{T_{1}TK}=A\frac{\Gamma }{\omega_{\rm c}^{2}+\omega_{\rm N}^2}$
 where $A$ depends on the hyperfine coupling constant and $K$
is the isotropic NMR shift.
Here, $\omega _{\rm c}$ corresponds to the fluctuation frequency of the
fluctuating hyperfine field at the $^{7}$Li
nucleus site transfered from the fluctuating Rh$^{4+}$ moments.
One would expect $R \approx  1/\omega_{\rm c}$ when 
$\omega_{\rm c}\gg\omega_{\rm N}$, while for
$\omega_{\rm c}\ll\omega_{\rm N}$ one should find that $R$
depends on the NMR field ($R\approx 1/\omega_{\rm N}$). 
When $\omega_{\rm c}=\omega_{\rm N}$, $R(T)$
approaches a maximum (see Supplemental Materials \cite{SM}, Fig.~S2),
which is a consequence of the slowing down of the fluctuation
frequency $\omega_{\rm c}$ of Rh$^{4+}$ moments. We find that
$\omega_{\rm c}$ is nearly $T$-independent at high-$T$, but decreases
below 10~K as $\omega_{\rm c}\propto T^{1.2}$ at low-$T$,
suggesting the slow spin dynamics of Rh$^{4+}$. This
is consistent with the
broad NMR line at low-$T$. The slowing down of spin fluctuations
might then dominate the low temperature magnetic properties \cite{muSR,YQM}. The
power law dependence of $1/T_1$ can be compared with that
found in the SOC-driven 5d-spin liquid compound
Na$_{4}$Ir$_{3}$O$_{8}$ \cite{ACS} and other low dimensional quantum magnets \cite{YS,TI2,XY,JY,PMRev}, which is attributed to the existence of gapless state in the spin
excitation spectrum and is in accord with finite value of $\chi$ and
$K$,
and \textit{C}$_{m}\sim $\textit{T}$^{2}$ behavior at low-$T$. 
The longitudinal nuclear spin-lattice relaxation rate is given by the low
energy ($\omega$) and momentum space ($q$)  integrated hyperfine form
factor $A(q,\omega)$ and the imaginary part of the
complex dynamic electron susceptibility $\chi^{\prime\prime}(q,\omega)$
(proportional to
$S(q,\omega)$, the
dynamic structure factor). For a 2D Kitaev spin
liquid calculations of $S(q,\omega)$ suggest either a gapped
($1/T_1\sim\exp(-\Delta/k_{\rm B}T)$) or a gapless
($1/T_1\sim T^n$) behavior \cite{RM1,JY,XY}. For Li$_{2}$RhO$_{3}$,
the gapless \textit{T}$^{2.2}$ power law or alternatively the
(pseudo) gapped behavior (i.e., $1/T_1\sim\exp(-\Delta/k_{\rm
  B}T)+\mbox{constant}$)  reasonably fit the $1/T_1$ data (see Supplemental Material 
\cite{SM}, Fig.~S4).

\begin{figure}
\includegraphics[width=\columnwidth]{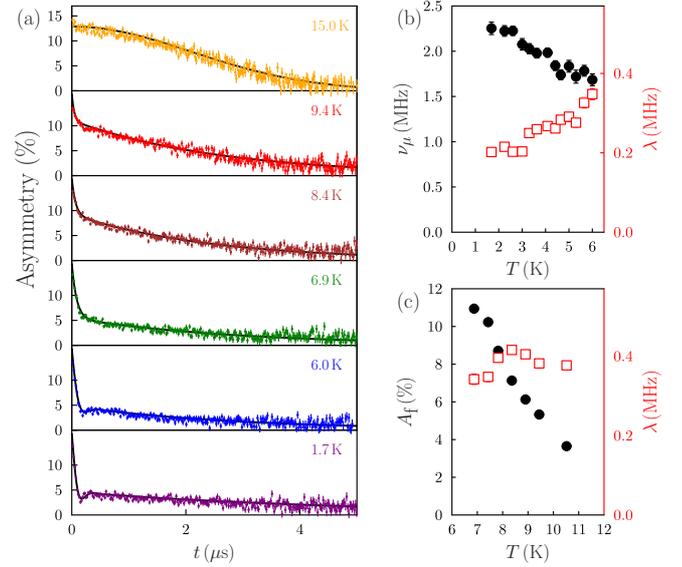}
\caption{(Color online) (a) Representative 
$\protect\mu $SR spectra for Li$_{2}$RhO$_{3}$ measured at a range of
temperatures. (b) The fitted precession frequency $\protect\nu _{\protect\mu %
}$ which indicates the average field experienced by the muon in the spin
frozen state (solid black circles, left-hand axis) and the relaxation rate $%
\protect\lambda $ of the slowly-relaxing component (open red squares,
right-hand axis). (c) The amplitude $A_{\mathrm{f}}$ of the fast-relaxing
component (solid black circles, left-hand axis) and $\protect\lambda $ (open
red squares, right-hand axis). }
\end{figure}
Our muon-spin rotation ($\mu$SR) \cite{musr}
experiments were carried out at PSI. Representative spectra are shown
in Fig. 4(a). For temperatures below about 6~K, there is a single
heavily damped oscillatory signal (proportional to $\cos (2\pi \nu _{\mu }t)%
\mathrm{e}^{-\Lambda t}$) together with a slow relaxation
(proportional to
$\mathrm{e}^{-\lambda t}$). The damped oscillation signifies static magnetic
Rh$^{4+}$ moments but the large damping is entirely consistent with
moment freezing, and is not associated with long range magnetic
ordering.
Fluctuations persist at low temperature evidenced by the presence of the
slow relaxation.
The
frequency of the damped oscillation, $\nu _{\mu }$, is around 2~MHz
and falls slightly on warming [see Fig.~4(b)] while the relaxation
rate
$\lambda $ rises. At \textit{T} above $\approx$6 K, it is no
longer possible to fit the fast relaxation with a damped oscillation, and we
identify this \textit{T} with the freezing temperature
$T_{\mathrm{g}}$,
in agreement with magnetization measurements \cite{YL,IIM} and coinciding
with the peak measured in \textit{ac} susceptibility (see Supplemental
Material~\cite{SM}).
For $T>T_{\mathrm{g}}$ we fit our data instead using a sum of two
exponential relaxations, so that the fitting function becomes
$A(t)=A(0)[A_{\mathrm{f}}\mathrm{e}^{-\Lambda t}+(1-A_{\mathrm{f}})\mathrm{e}^{-\lambda
 t}] $.
The amplitude of the fast relaxing term $A_{\mathrm{f}}$ falls on
warming above 6~K [see Fig.~4(c)] and is entirely absent by
15~K by which temperature the relaxation is dominated by a
Gaussian response characteristic of static nuclear moments. These
observations indicate that although the freezing disappears above
$T_{\mathrm{g}}$ there remain some frozen regions of the sample which persist
well above $T_{\mathrm{g}}$, up to almost $2T_{\mathrm{g}}$, perhaps in
small, slowly-fluctuating clusters. These slow fluctuations may likely
contribute to the slow relaxation that is observed in these data. The volume
fraction of the clusters decreases on warming and this can be directly
related to the decrease in $A_{\mathrm{f}}$. In fact, the observation of a
slowly relaxing fraction throughout the temperature range demonstrates that
the frozen state possesses some weak dynamics. These measurements are
consistent with the development of a moment-frozen state below 6~K
at small magnetic fields.

\section{Conclusion}

We have presented a study on the magnetism of Li$_{2}$RhO$_{3}$ to
probe a possible spin liquid ground state and investigate the presence
of partial frozen moments. Whereas frozen moments were evidenced by
low field susceptibility measurements, the NMR measurements performed
in higher field could hardly resolve this effect. The zero field
$\mu$SR evidences frozen moments but also persistent low energy spin
dynamics.  $\chi $, $\delta H$ and $K$ remain finite towards low $T$
wheras the magnetic specific heat as well as the spin lattice
relaxation rate exhibit characteristic temperature dependencies
assigned to quantum spin liquids.  The magnetic heat capacity $C_{\rm
  m}$ displays no signature of LRO down to 0.35~K despite an AFM
interaction $J/k_{\rm B}\approx 75$~K between Rh$^{4+}$ moments. The
$C_{\rm m}\sim T^{2.2}$ and the $1/T_1\sim T^{2.2}$ behavior at
low-$T$ might be assigned to gapless excitations as predicted for the
Kitaev quantum spin liquid state. Further studies on single crystals
are highly recommended to establish whether the partial moment
freezing is generic to the system (e.g.\ because of the proximity of
the Kitaev QSL to the magnetic ordered phase) or a matter of sample
quality (presence of structural disorder). Nonetheless the study
presented here clearly show that Li$_{2}$RhO$_{3}$ is not a
conventional bulk spin glass material, the SRO effects are wiped out
in magnetic fields and most important the low-$T$ spin dynamics as
well as the specific heat are reminiscent of that of a quantum spin
liquid.

\section{Acknowledgement}

 We would like to thank A. V. Mahajan, H. Yasuoka,
P. Mendels and M. Majumder for fruitful discussions. PK acknowledges support
from the European Commission through Marie Curie International Incoming
Fellowship (PIIF-GA-2013-627322). SM acknowledges financial support by the
Helmholtz Virtual Institute 521 (\textquotedblleft New states of matter and
their excitations\textquotedblright ). FRF, TL and SJB acknowledge support
from EPSRC (UK) (Grants EP/M020517/1 and EP/N023803/1).
Part of this work was carried out at the Swiss Muon Source
(S$\mu$S), Paul Scherrer Institut, Villigen, Switzerland.

\end{document}


\preprint{APS/123-QED}
\title{\textbf{Supplemental Material}: Local Magnetism and Spin Dynamics in
the Honeycomb Rhodate Li$_{2}$RhO$_{3}$}
\date{\today}
\author{P. Khuntia}
\altaffiliation[Present address: ]{Department of Physics, Indian Institute of Technology, Madras Chennai-600036, India}
\affiliation{Max Planck Institute for Chemical Physics of Solids, 01187 Dresden, Germany}
\author{S. Manni}
\altaffiliation[Present address: ]{Department of Physics and Astronomy, Iowa State University, Ames, Iowa, USA}
\affiliation{EP VI, Center for Electronic Correlations and Magnetism, Augsburg
University, D-86159 Augsburg, Germany}
\author{F. R. Foronda}
\affiliation{Oxford University Department of Physics, Clarendon Laboratory,\\
Parks Road, Oxford OX1 3PU, United Kingdom}
\author{T. Lancaster}
\affiliation{Durham University, Centre for Materials Physics, South Road, Durham, DH1
3LE, United Kingdom}
\author{S. J. Blundell}
\affiliation{Oxford University Department of Physics, Clarendon Laboratory,\\
Parks Road, Oxford OX1 3PU, United Kingdom}
\author{P. Gegenwart}
\affiliation{EP VI, Center for Electronic Correlations and Magnetism, Augsburg
University, D-86159 Augsburg, Germany}
\author{M. Baenitz}
\thanks {corresponding author: baenitz@cpfs.mpg.de}
\affiliation{Max Planck Institute for Chemical Physics of Solids, 01187 Dresden, Germany}

\begin{abstract}
\end{abstract}
\maketitle

        \setcounter{figure}{0}
        \renewcommand{\thefigure}{S\arabic{figure}}

\section{Sample synthesis, magnetic exchange and effect of lattice site disorder}
We have synthesized Li$_{2}$RhO$_{3}$ polycrystals by the solid state
reaction method from stoichiometric amounts of  Li$_{2}$CO$_{3}$
and Rh-powder. The mixture has been pelletized several times 
and reacted in  a flow of O$_{2}$ at temperatures up to 850$^\circ$C. Powder
x-ray-diffraction (XRD) scans do not reveal any evidence for
secondary phases and are similar to those reported in  Refs.~\onlinecite{VT,SMT}.
There is no evidence for site mixing between Li and Rh atoms. However, as 
found for Na$_{2}$IrO$_{3}$ \cite{SKC}, frequent in-plane
translational stacking faults along the $c$-axis are probably
present in our sample of Li$_{2}$RhO$_{3}$. 
Nevertheless, this is not likely to affect the in-plane 2D physics.
Edge-sharing RhO$_{6}$ octahedra form a Rh-honeycomb arrangement in
which each Rh ion carries a spin-orbit entangled $j_{%
\mathrm{eff}}=\frac{1}{2}$ moment. The lattice structure hosts three different Li-sites (one within and two out of the honeycomb planes \cite{SMT}).

\section{dc- and ac- susceptibility}
The d.c.\ magnetic susceptibility ($\chi
$=\textit{M}/\textit{H}) was measured using a SQUID magnetometer
(MPMS, Quantum Design, QD) in the temperature\textit{\ }range 1.8
$\leq T\leq $ 300 K in small magnetic fields ($\mu_0H\leq $1~T) where
partial spin freezing was found and at 4.23~T and 6.89~T at which
NMR\ measurments were carried out. The frequency and temperature
dependence of the a.c.\ susceptibility data were obtained using a
commercial a.c.\ susceptometer (PPMS, Quantum Design). We found a
pronounced peak at $\textit{T}_g$= 6 K in the
a.c.\ suceptibility indicative of partial spin
freezing, accompanied by a weak frquency dependence reminiscent of a
spin glass (see Fig.~S1).  The effective exchange coupling between
Rh$^{4+}$ using mean field theory (MFT) can be estimated as: $J/k_{\rm
  B}=-3\theta _{CW}/zS(S+1)=$80~K (with coordination
number $z=3$). This is a very simple and crude estimate
given the presence of various exchange couplings in Li$_{2}$RhO$_{3}$
of comparable strengths \cite{II,SKC}.

\begin{figure}
\includegraphics[width=\columnwidth]{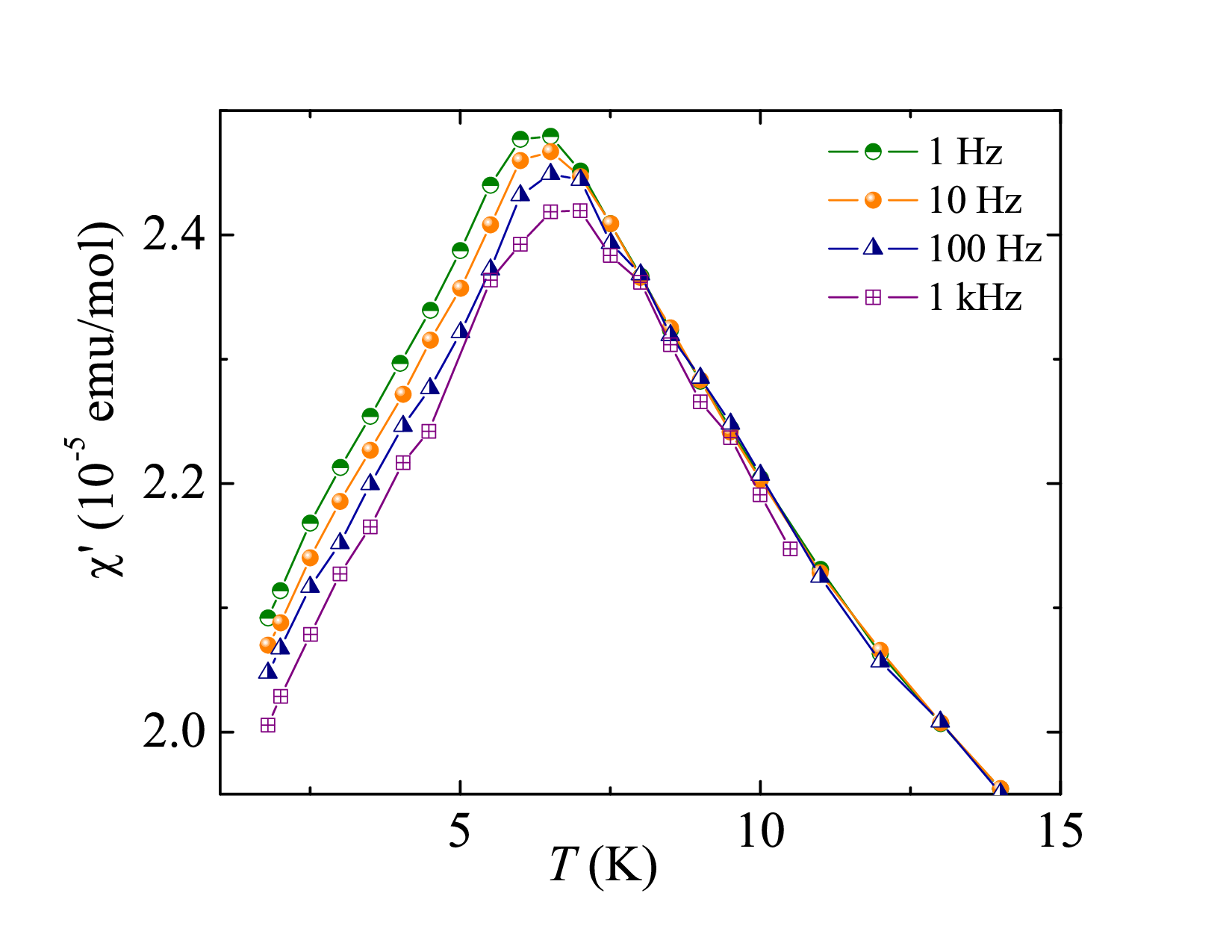}
\caption{(Color online) The temperature
dependence of the real part of the a.c.\ magnetic susceptibility at various
frequencies.}
\end{figure}

For many $S=\frac{1}{2}$ quantum magnets the exchange interaction
between the nearest neighbor spins can be
described by a Heisenberg type hamiltonian,
$H = J\sum_{<i,j>} S_{i}\cdot S_{j}$.  The magnetic susceptibility at high temperatures can be modelled by the high temperature series expansion (HTSE). 
For pseudospin $s=\frac{1}{2}$ Rh$^{4+}$ ions with moderate SOC on the
honeycomb lattice the model is of some use (especially if the exchange
is predominant antiferromagnetic and the Kitaev term is small) and the
susceptibility is then given by
\begin{equation}
\chi =\frac{N_{A}g^{2}\mu _{B}^{2}}{4k_{B}T}\sum (\frac{J}{k_{B}T})^{^{n}},
\end{equation}
where $n = 0.5$ and $a_{n}$ are series coefficients, the values of
which can be found in Ref.~\onlinecite{HTSE2}.  We obtain $J/k_{\rm
  B}\approx (75\pm 5)$~K, which is close to that estimated using
MFT. Given the simplicity of the model and appreciable spin orbit
interactions with $J_{\rm eff}=\frac{1}{2}$, the obtained
  coupling constant is a rough estimate.  It may be noted that the
  HTSE is strictly valid only for $S=\frac{1}{2}$ systems in the
  absence of spin orbit interactions.

\section{$^{\mathbf{7}}$Li NMR:
linewidth analysis and spin-lattice relaxation}

NMR is a useful local probe for spin dynamics and local disorder. The local
disorder (due to defects, vacancies or anti site order) usually shows
in an enhanced NMR linewidth $\delta H$, whereas the dynamical
information arises via
the spin lattice relaxation rate which reflects the fluctuation
of hyperfine fields. The NMR spectra and spin-lattice relaxation
measurements at two frequencies were carried out using a standard pulsed NMR
spectrometer.

\begin{figure}
\includegraphics[width=\columnwidth]{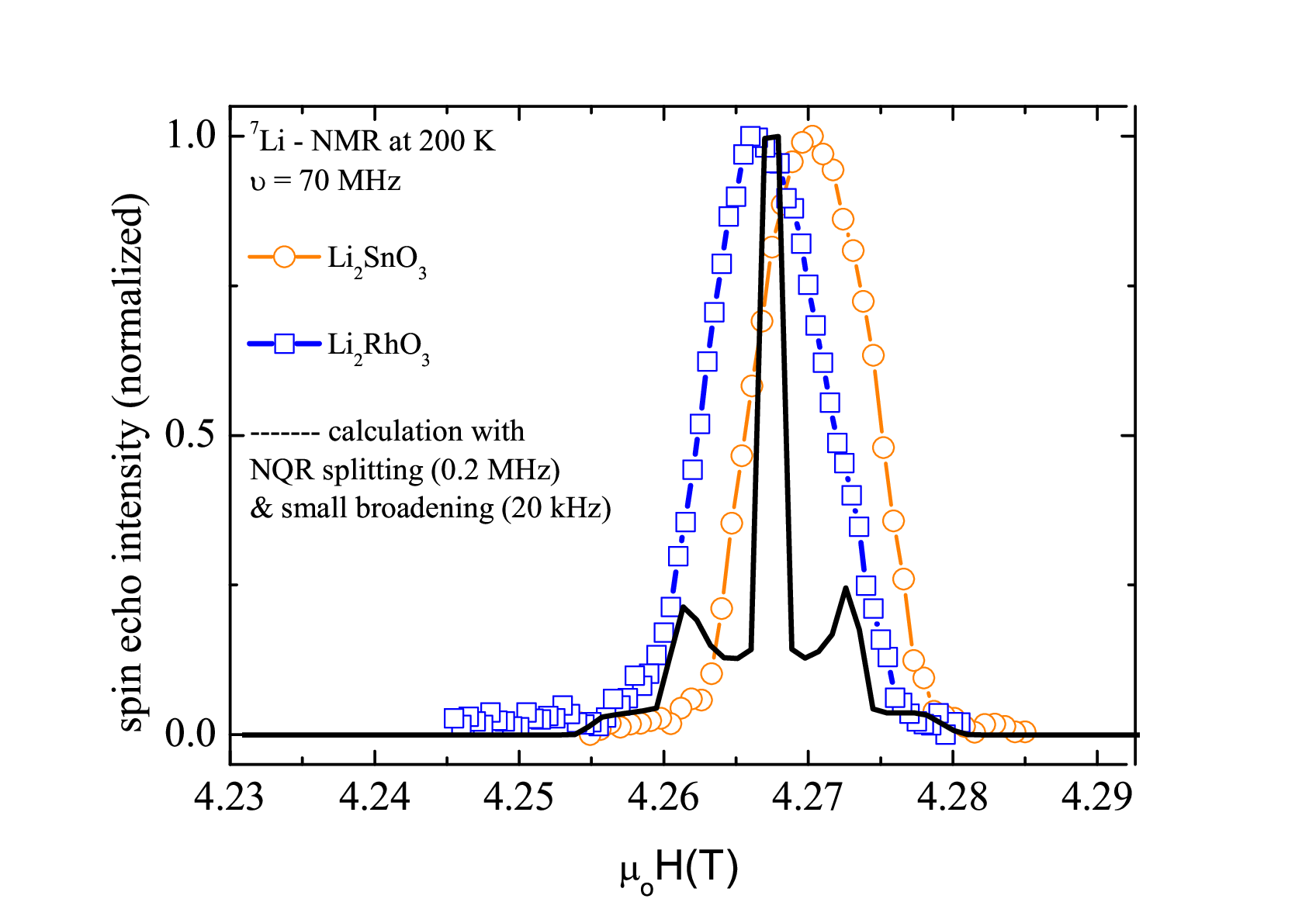}
\caption{(Color online) $^{7}$Li -powder
spectra at 200~K for Li$_{2}$RhO$_{3}$ and the non magnetic homologue
compound Li$_{2}$SnO$_{3}$. In addition a single Li-site calculation of the
spectra with negligibly small broadening is shown by the solid line.}
\end{figure}

The \textit{T}-dependence of $\delta H$ in Li$_{2}$RhO$_{3}$ was
compared with that measured in the non-magnetic reference
Li$_{2}$SnO$_{3}$. This allows the investigation of
the effect of the quadrupolar broadening
on the spectra. The linewidth of Li$_{2}$RhO$_{3}$
($\delta H=\delta H_{\rm 4d}+\delta H_Q$) has a $T$-dependent
contribution ($\delta H_{\rm 4d}$) related to the underlying
Rh magnetism,  plus a (nearly) $T$-independent
contribution ($\delta H_Q$) arising
from the (powder averaged) quadrupolar transitions
of the $I=\frac{3}{2}$ quadrupolar $^{7}$Li nuclei.
For Li$_{2}$SnO$_{3}$ the linewidth originates only from the
quadrupole effect ($\delta H_{\rm 4d}=0$).  As seen in Fig.~3, the magnetic
brodening in Li$_{2}$RhO$_{3}$ becomes dominant below approximately
50~K (where
$\delta H_Q$ saturates) wheras
towards higher temperatures the linewidth mainly originates from the
$T$-dependence of the quadrupole contribution.

\begin{figure}
\includegraphics[width=\columnwidth]{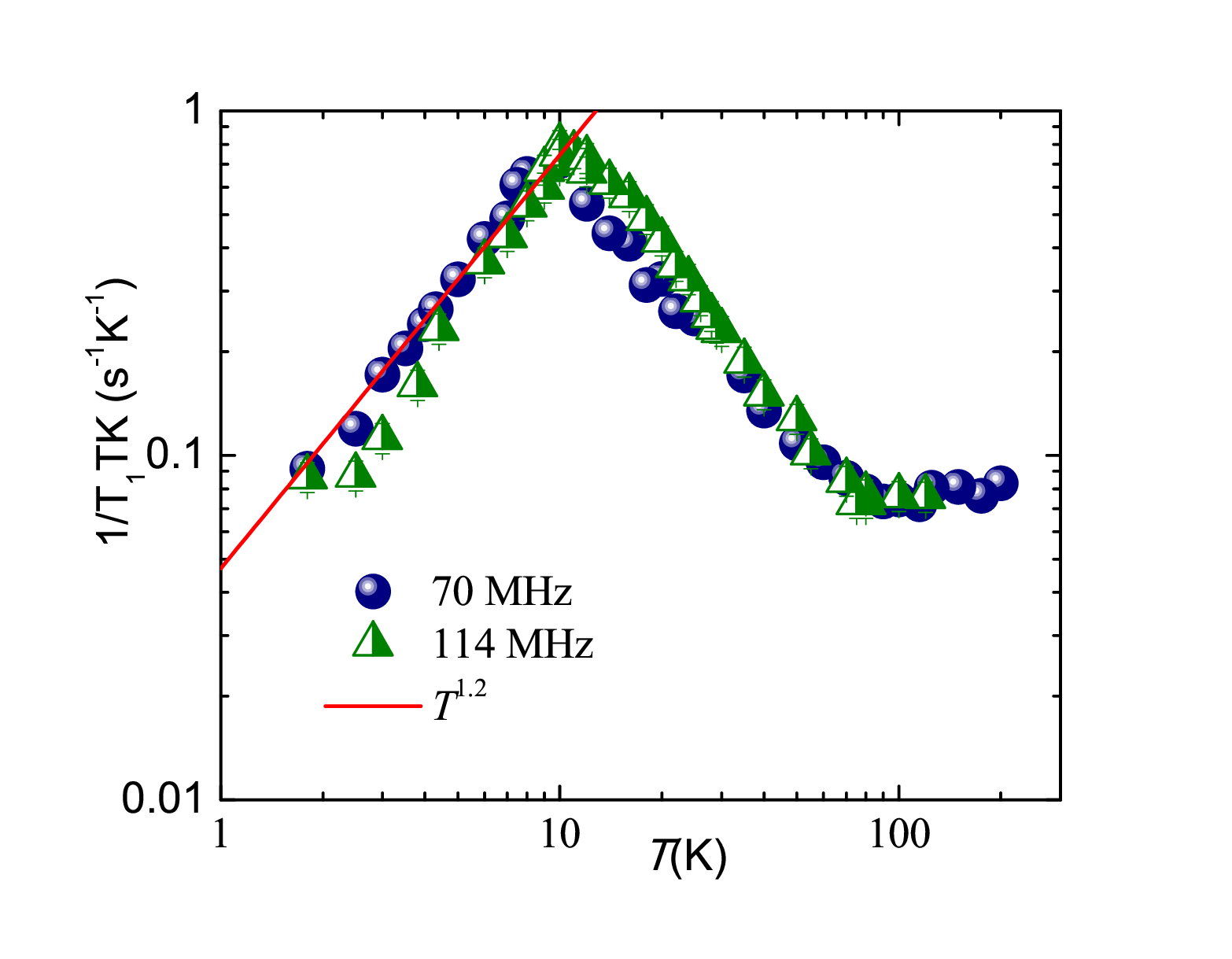}
\caption{(Color online) The temperature
dependence of $1/(T_{1}TK)$ at two NMR frequencies.}
\end{figure}

To illustrate this we have shown the $^7$Li-NMR lines at 200~K for both
compound together in Fig.~S2. In addition a model calculation for the
quadrupolar-split NMR line is shown. Both lines are of equal width, but
one should note that there are three different
Li positions in the structure (one within and two out of the honeycomb
planes
\cite{SMT}) which also (beside the powder
average) lead to some averaging of the total spectra. Having this in mind it
is quite surprising that both lines are relatively narrow and exhibit nearly
no spectral anisotropy. The equal linewidth at high temperatures also
gives evidence for the absence of anti-site order
(Li-Rh and Li-Sn site mixing) frequently discussed in the literature.

\begin{figure}
\includegraphics[width=\columnwidth]{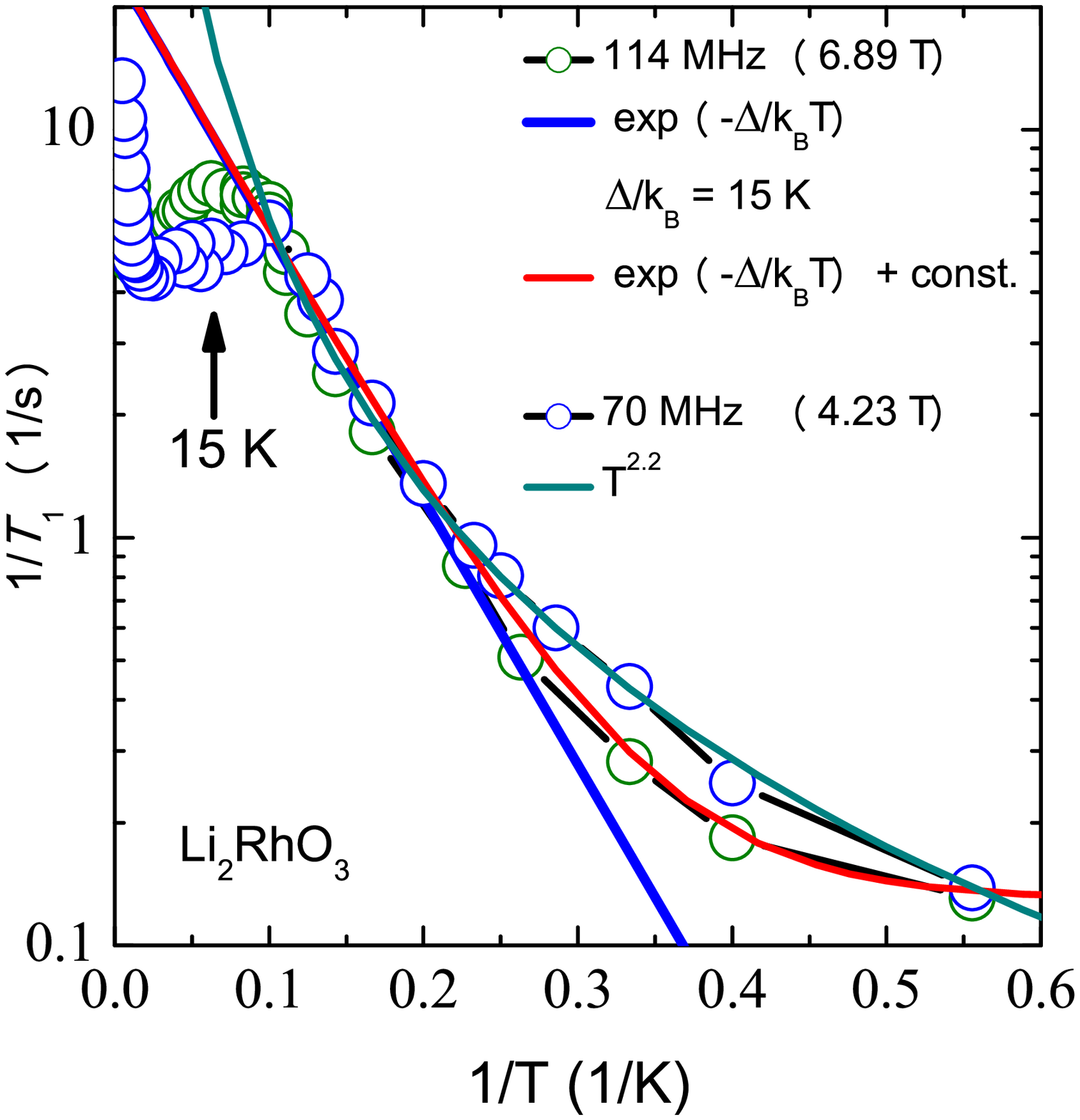}
\caption{(Color
online) Arrhenius plot for the NMR relaxation rate for the two NMR
frequencies together with different fitting functions (see text).}
\end{figure}
The spin lattice relaxation measurements
were performed following the saturation recovery method using short \textit{%
rf} pulses. In Li$_{2}$RhO$_{3}$, the recovery of longitudinal nuclear
magnetization, $M(t)$, at time $t$ after the
saturation pulse could be fitted with
$1-M(t)/M(\infty)={\rm e}^{-(t/T_{1})^{\beta }}$ (here $M(\infty)$ is the
saturation magnetization, and $\beta$ ($\approx 0.5$) is the stretch
exponent, which is found to be independent of temperature here).  This
function arises when there is a distribution of relaxation times.
The spin-lattice
relaxation rate, $1/T_{1}$ at each temperature is determined by fitting the
nuclear magnetization $M(t)$ using the stretched exponential function (see
Fig.~5(b)). A non-exponential autocorrelation function is frequently found
in quantum spin liquids, but is also common in disordered materials. In
general, $1/T_1$ probes the $q$-averaged low energy spin
excitations and $1/(T_1T)$ can be expressed as the wave-vector 
$q$-summation of the imaginary part of the dynamic electron spin
susceptibility $\chi ^{\prime \prime }(q,\omega_{n})$:
$\frac{1}{T_{1}T}\propto \sum_{q}\mid
A_{hf}(q)\mid ^{2}\frac{\chi ^{\prime \prime
  }(q,\omega _{n})}{\omega _{n}}$.
Here, $A_{hf}(q)$ is the form factor of the
hyperfine interactions and $\omega_{n}$ is the nuclear Larmor frequency 
\cite{TM2}. \ Fig.~S3 shows the temperature dependence of
$R(T)=1/(T_1TK)$
as a sort of "on site" fluctuation rate.
Above 100~K, $R(T)$ is constant which indicates local moment
magnetism, consistent with the Curie-Weiss behaviour of the
susceptibility.
However, towards low temperatures strong correlations between Rh moments set
in which is reflected by an increase of $R(T)$. Around 10~K these correlations
start dying out towards low temperatures which indicates
a slowing down of spin fluctuations. 
One way to discuss whether the ground state is spin gapped or gappless is the Arrhenius
type of plot (see Fig.~S4). Here the experimental data points from
Fig.~5 in the main
text are replotted together with the $T^{2.2}$ power law and the
simple equation for excitations across a gap without
($1/T_1\sim \exp(-\Delta /k _{B}T)$, shown by the straight line in Fig.~S4) and with a residual
(constant) value. A residual $1/T_1$
value might originate from a weak additional Heisenberg type of
exchange coupling. Furthermore, a sizable gap anisotropy might also
lead to a residual value of $T_1$.
 Within the experimental accuracy, we cannot
distinguish between a $T^{2.2}$ power law and a pseudogapped behaviour
(including a residual value of the relaxation rate) in the
spin-lattice relaxation rate.  Therefore, our conclusion is that our
NMR measurements are probing
gapless excitations, consistent with the
specific heat results.

\section{Muon-spin rotation}
The muon-spin rotation ($\mu$SR) measurments on the polycrystalline
sample of Li$_{2}$RhO$_{3}$ shown in Fig.~6 of the main paper were
carried out on the general-purpose spectrometer (GPS) at the Swiss
Muon Source (S$\mu $S) at PSI (Switzerland).  The $\mu$SR data were
analyzed using the analysis software WiMDA \cite{wimda}.  In a $\mu$SR
experiment spin-polarized muons are implanted into a sample, where
they Larmor precess around the local magnetic field at the muon
stopping site. By measuring the angular distribution of the decay
product positrons the spin polarization can be tracked. In the case of
long-range magnetic order, coherent magnetic fields at particular muon
stopping sites within the unit cell lead to oscillatory signals with
frequencies dependent on the local magnetic fields at each site. In
$\mu$SR, impurity phases only contribute according to their volume
fraction, and so the technique is an effective measure of intrinsic
behavior.